# The plasma protein fibrinogen stabilizes clusters of red blood cells in microcapillary flows

M. Brust[1,2*], O. Aouane[1,2,6*], M. Thiébaud[2*], D. Flormann[1], C. Verdier[2], L. Kaestner[3], M. W. Laschke[4], H. Selmi[5,6], A. Benyoussef[7], T. Podgorski[2], G. Coupier[2], C. Misbah[2], C. Wagner[1]

1 Experimental Physics, Saarland University, 66123 Saarbrücken, Germany
2 Laboratoire Interdisciplinaire de Physique, CNRS - UMR 5588, Université Grenoble I, B.P. 87, 38402 Saint Martin d'Hères Cedex, France
3 Institute for Molecular Cell Biology and Research Centre for Molecular Imaging and Screening, School of Medicine, Saarland University, Building 61, 66421 Homburg/Saar, Germany
4 Institute for Clinical & Experimental Surgery, Saarland University, 66421 Homburg/Saar, Germany
5 Laboratoire d'Ingénierie Mathématique, Ecole Polytechnique de Tunisie B.P. 743 - 2078 La Marsa, Tunisia
6 Riyadh College of Technology, Technical and Vocational Training Corporation, Riyadh 12433, Saudi Arabia
7 LMPHE, URAC 12, Faculté des Sciences, Université Mohammed V- Agdal, Rabat, Morocco
*These authors contributed equally to this work

Correspondence regarding the experimental work should be addressed to C.W. (c.wagner@mx.uni-saarland.de), and correspondence regarding the numerical work should be addressed to C.M. (chaouqi.misbah@ujf-grenoble.fr).

**Abstract**

**The supply of oxygen and nutrients and the disposal of metabolic waste in the organs depend strongly on how blood, especially red blood cells, flow through the microvascular network. Macromolecular plasma proteins such as fibrinogen cause red blood cells to form large aggregates, called rouleaux, which are usually assumed to be disaggregated in the circulation due to the shear forces present in bulk flow. This leads to the assumption that rouleaux formation is only relevant in the venule network and in arterioles at low shear rates or stasis. Thanks to an excellent agreement between combined experimental and numerical approaches, we show that despite the large shear rates present in microcapillaries, the presence of either fibrinogen or the synthetic polymer dextran leads to an enhanced formation of robust clusters of red blood cells, even at haematocrits as low as 1%. Robust aggregates are shown to exist in microcapillaries even for fibrinogen concentrations within the healthy physiological range. These persistent aggregates should strongly affect cell distribution and blood perfusion in the microvasculature, with putative implications for blood disorders even within apparently asymptomatic subjects.**

**Introduction**

The physical mechanism of the plasma protein-induced aggregation process is controversial[1-5]. Although the first studies favoured a model based on bridging[6-9], a model based on depletion was also introduced [10-13]. The strength of aggregation depends on the physical properties of red blood cells (RBCs), such as deformability[14,15] and surface charge[16]. In blood plasma, fibrinogen has been identified already in the 1960´s as the main protein responsible for aggregation[7,17-19]. In pathological cases such as cardiac diseases[20] or sepsis[21], both an increased level of fibrinogen and an accumulation of RBC aggregates with more robust clusters have been found. In addition to fibrinogen, model polymers such as dextran can be used to induce aggregation. Recently, the interaction strength between two RBCs was determined at various dextran concentrations using single cell force spectroscopy[13]. An increase in dextran concentration has been found to initially induce an increase in the interaction strength among RBCs, which is abolished at high concentrations.

Because the viscosity of blood from healthy donors exhibits strong shear thinning up to shear rates of approximately 100 s$^{-1}$, when all aggregates are broken up[3,22], one could expect aggregation to be negligible in the microcirculation where average shear rates are higher. Nevertheless, about 80% of the total pressure drop in blood circulation occurs in the vascular network[3] that irrigates organs in which oxygen and nutrients are exchanged, and contradictory results have been reported on the effect of aggregation in glass capillaries of 30 to 130 µm diameter[23,24] and *in vivo* vascular networks; both higher and lower flow resistances are reported after the addition of aggregating agents[22]. Two effects appear to counteract each other: First, the aggregated cells appear to align more strongly along the centre line, thus decreasing the flow resistance; however, blocking might occur at the entrance of the capillary network, thereby increasing the pressure drop. Interestingly, athletic mammalian species exhibit higher RBC aggregability[1,25], which raises the question whether aggregation might indeed be a way to enhance RBC throughput in the circulation and thus improve the oxygen supply in the body. Theoretical descriptions of the flow of blood or suspensions of vesicles in microcapillaries have improved considerably in recent years. For instance, numerical studies have shown that even very weak hydrodynamic interactions are sufficient to stabilise clusters in capillaries[26,27] The viscosity of blood has been numerically calculated[28], and a comparison of this value with experimental rheological data led to a predicted value for the effective interaction force between two RBCs of approximately 5 pN, a value that is considerably lower than that found here.

**Results**

Intravital microscopy observations of the microcirculation in mice (Fig. 1a, a movie is available in the online supplementary material) show that RBCs can indeed flow as isolated cells or as clusters in capillaries, but the relative influence of hydrodynamic and aggregation forces remains unclear. We carried out in vitro investigations (Fig. 1b) and numerical simulations (Fig. 1c-d) to quantify this phenomenon. In the experiments, RBCs at a very low haematocrit (less than 1%) are pumped through a microfluidic device. At a typical flow velocity of v = 1 mm/s, which is comparable to physiological values and corresponds to a wall shear rate of approximately 500 $s^{-1}$, we find that the number of clusters strongly depends on the dextran or fibrinogen concentration. A bell-shaped dependency curve of the population of clusters with two or more cells for dextran (Fig 2a) and a monotonic increase for fibrinogen (Fig 2b) are observed. Remarkably, robust clusters form within the range of healthy human physiological conditions of fibrinogen (1.8 - 4 mg/ml[29]).

In numerical simulations of interacting RBCs in capillaries, we considered a two dimensional geometry, which has proven to capture many of the behaviours of RBCs in three dimensions[27]. The cell model comprises closed inextensible membranes (giant vesicles) filled with fluid (modelling haemoglobin) and suspended in a different fluid (modelling plasma). The vesicles can freely deform in response to the imposed Poiseuille-type flow. Cells can interact by both hydrodynamic forces and by an additional interaction potential with interaction energies that were taken from single cell force spectroscopy measurements with fibrinogen. We take an initial set of five cells in two channels of widths 4.5 and 12 µm, which are typical values in human capillaries (4.5 µm corresponds to the height of our experimental channels and 12 mm where we found that the clusters persist even though they could experience stronger shear stresses out of the center and pass on top of each other). The velocities explored at the centre of the channel are approximately 1 mm/s. Cells are disposed either randomly in the channel or in a line. We progressively increase the interaction energy and analyse the subsequent behaviour of the initial configuration. Fig. 1c-d summarises our findings. The interaction energy is increased from top to bottom in the figure. At low interaction energies, the cells that are initially arranged in a cluster are quickly separated by the flow. Interestingly, when the energy reaches the range corresponding to physiological concentrations, small clusters of two (and occasionally three) cells begin to form and persist. When the energy is such that the concentration approaches the maximal physiological concentration, five cells form a stable cluster. The evolution of the stable cluster size vs. interaction energy is summarised in Fig. 2c-d. Both the experimental and the numerical results show clustering transitions at the same interaction energies.

To measure the interaction energies between two discocytic RBCs in a rouleau at rest (Fig 3a-b), we used single cell force spectroscopy with an atomic force microscope. A RBC was attached to a cantilever and brought into contact with another RBC resting on the cover slip; the work needed to separate the two RBCs was then measured at various dextran and fibrinogen concentrations (Fig. 3c). The interaction energy was then determined from the force-distance curve[13]. For small and large dextran concentrations, the measured interaction energy goes to zero; a maximum is found at intermediate concentrations. For fibrinogen, the interaction energy is found to increase monotonically. This dependency is possibly explained by a model based on depletion forces[30]. This means that there is a depletion layer between the adhering cells with no macromolecules and the surrounding macromolecules cause an osmotic pressure that pushes the two cells together. It is furthermore assumed that the dextran can penetrate the glycocalyx at higher concentrations, thus cancelling the effect of depletion. Due to its negative charges, fibrinogen is assumed to remain outside the glycocalyx layer and thus the interaction energy continuously increases with the concentration.

The shape of the interaction energy vs. concentration curves (Fig. 3c) is remarkably consistent with the statistics of cluster sizes in the microfluidic experiments (Fig. 2a-b). The same behaviour was found when measuring the erythrocyte sedimentation rate (Fig. 3d), a quantity that is largely used in blood diagnostics to characterise RBC aggregation and indicate inflammatory states.

This ensemble of results leads to the conclusion that the relatively weak interaction forces between cells do indeed have a strong impact on blood flow in the microvasculature by enhancing clustering, regardless of the high wall shear rates present in our experiments and in capillaries in vivo, which were suspected to prevent the formation of clusters. Indeed, the configuration of RBCs in capillaries decreases the extensional stresses between cells that tend to separate them from one another in bulk shear flow. To further stress this major difference with the macrocirculation, we conducted further experimental investigations using classical shear rheometry. The well-known shear-thinning behaviour of blood due to the destruction of aggregates is recovered here with RBC suspensions in the presence of various concentrations of dextran (Fig. 4a). When computing the relative viscosity (the ratio of the viscosity of the RBC suspensions to that of the suspending medium) to extract the effect of the aggregates on the viscosity (Fig. 4c), the bell-shaped dependency of the viscosity is revealed. Again, this is different for fibrinogen, which is charged (Fig. 4b,d). For fibrinogen, the relative viscosity monotonically increases with concentration. However, for both macromolecules, a negligible effect of aggregation is found at higher shear rates (greater than 10 $s^{-1}$ for fibrinogen and greater than 50 $s^{-1}$ for dextran), in a marked contrast with the present microfluidic experiments where shear rates are as high as 500 $s^{-1}$. Indeed, in the situation of highly confined microfluidic channel flow, the red blood cells are kept in the middle of the Poiseuille profile and therefore experience lower overall hydrodynamic stresses compared to the bulk rheology where clusters are exposed to simple shear that sufficiently stretches them to break them at equivalent shear rates.

**Discussion**

A microfluidic device was used to study the dextran and fibrinogen induced aggregation of red blood cells in microcapillaries. We used quantitative data from single cell force spectroscopy on the interactions energies of fibrinogen induced aggregates of red blood cells to perform numerical simulations of the cluster formation. The results are in excellent agreement with the experimental results. This finding suggests that robust clusters can survive the relatively high shear stresses experienced in the microvasculature and they might survive over quite a long distance in the microvasculature, despite the large shear rates. This should affect not only the local rheology of blood but also the RBC distribution in the microcirculatory network and the related oxygen delivery to tissues through the potential impact of clusters on cell distribution at bifurcations. Cluster formation should strongly influence the entrance of RBCs into and their distribution within capillaries and their effect on oxygen delivery. RBC aggregation is generally known to increase microvascular flow resistance[31] and consequently reduce blood perfusion to the organs[32]. An extended lack of perfusion over time causes capillaries to become dysfunctional, a first stage towards ischemia. Patients suffering ischemic stroke exhibit high concentrations of fibrinogen (3.8-6.2 mg/ml)[23]. Understanding the long-term consequences of persistent cluster formation is therefore crucial to propose new predictive markers and treatment strategies.

**Materials and Methods**

*Imaging of RBCs in vivo*

Animal experiments were approved by the governmental animal care committee of the Saarland, Germany, and were conducted in accordance with German legislation on the protection of animals and NIH Guidelines for the Care and Use of Laboratory Animals (NIH Publication #85-23 Rev. 1985).

Circulating RBCs were imaged in the dorsal skinfold chamber of BALB/c mice (22-25 g) using intravital fluorescence microscopy. The dorsal skinfold chamber and its implantation procedure have been described previously in detail[33]. After surgery, the mice were allowed to recover for 72 h. Subsequently, the animals were anaesthetised by i.p. injection of ketamine (75 mg/kg body weight; Ursotamin®; Serumwerk Bernburg, Bernburg, Germany) and xylazine (15 mg/kg body weight; Rompun®; Bayer, Leverkusen, Germany) and fixed in the right lateral decubital position on a Plexiglas stage. For the visualisation of microvessels by the contrast enhancement of blood plasma, 0.05 ml 5% fluorescein-isothiocyanate (FITC)-labelled dextran (molecular weight: 150,000 Da; Sigma-Aldrich, Taufkirchen, Germany) was injected i.v. via the retrobulbary space. The dorsal skinfold chamber was positioned under an upright microscope (E600; Nikon, Tokyo, Japan) equipped with a 100x, NA 1.2, water immersion objective and a halogen lamp attached to a FITC filterset. The microscopic images were recorded using a charge-coupled device video camera (iXon Ultra; Andor, Belfast, UK) connected to a PC system. For image processing, the black and white pictures were changed into red (original: black) and black (original: white) pictures using look-up tables.

*Sample preparation*

The study was approved by the ethics committee of the Medical Association of the Saarland (reference No 24/12). We obtained informed consent from the donors after the nature and possible consequences of the studies were explained.

Microfluidics and AFM measurements: A droplet of blood was collected via finger prick from a healthy donor and diluted in 1 ml of physiological buffer solution (phosphate buffered saline, PBS, Invitrogen, Darmstadt, Germany). To remove all substances other than red blood cells (RBCs), the sample was centrifuged at 3,000 rpm for 3 minutes, and the liquid phase and buffy coat were removed completely; the sample was then rediluted with 1 ml of PBS. To measure cluster size in microfluidic channels, 1 mg/ml of bovine serum albumin (BSA, Polysciences, Warrington, USA) was added to the buffer solution in this last step. In the AFM experiments, BSA was added after a cell had been glued to the cantilever. CellTak® (BD Biosciences, Franklin Lakes, USA), which was used for the functionalisation step, is passivated by the BSA. To prepare the solution with the depletion agent, either dextran (MW = 70 kDa) or fibrinogen (both in powder form, Sigma Aldrich, , Taufkirchen, Germany) was weighed into a test tube and diluted with 1 ml of PBS by gentle shaking. For the microfluidic experiments, the solution with the RBCs and the solution with the depletion agent were mixed immediately prior to filling of the syringe. For the measurements in the AFM experiment, the solution containing the depletion agent was added to the RBC solution after a cell had been glued to the cantilever. The experiments were performed within four hours of the blood collection.

Rheometer and sedimentation measurements: Venous blood of healthy donors was drawn into standard EDTA tubes. RBC enrichment was performed as described above. The obtained RBC concentrate was stored at 4°C and mixed immediately prior to the measurements with PBS containing the amount of fibrinogen or dextran required to reach a volume concentration (haematocrit) of 45%, a typical in vivo value. The experiments were performed on the same day as the blood collection.

*Rheometry*

Rheological measurements of blood cell suspensions in the dextran solutions were carried out using a commercial constant shear rheometer (Mars II, Thermo Scientific, Karlsruhe, Germany) equipped with a cone-plate geometry (diameter: 60 mm, angle: 1°) at a temperature of 23°C. Samples were pre-sheared at 100 $s^{-1}$ for one minute to homogenise the samples, limit sedimentation and obtain comparable initial conditions. Consequently, the viscous stress was determined at 21 different shear rates between 10 and 100 $s^{-1}$ (these values correspond to the regions where correct viscosities can be obtained). For each measurement point, we ensured that the cone was rotating first for at least one and a half revolutions; then, the torque was averaged over at least one revolution to obtain a good resolution at low shear rates. Measurements with fibrinogen were carried out on a low-shear Contraves LS-30 rheometer (prorheo GmbH, Althengstett, Germany) for greater sensitivity using a Couette geometry (outer radius 5.5 mm, gap 0.5 mm, height 20 mm). For each value of the shear rate, a steady value of the torque was obtained after a rotation period of 15 s.

*Single cell force spectroscopy*

An AFM (JPK Instruments, Berlin, Germany) was used in single cell force spectroscopy measurements[13] to determine the interaction energies between RBCs in the presence of dextran or fibrinogen. The AFM was equipped with a special piezometer (CellHesion. JPK Instruments, Berlin, Germany) that allowed large vertical displacements (up to 100 µm). The cantilevers used (MLCT-O, no tip, Bruker, Billerica, USA) had a very low spring constant (approximately 0.01 N/m), allowing for small force measurements with good accuracy (approximately 5 pN).

Fresh RBCs were obtained from healthy donors using a finger prick. Cells were washed three times by centrifugation (800 g, 3 min) in PBS as described above and allowed to rest in a Petri dish.

The cantilever was treated as follows: after a 2-min incubation in Cell-Tak (BD Biosciences, Franklin Lakes, USA), the cantilever was allowed to dry for 3 min; then, it was rinsed in ethanol and PBS. Cell-Tak was used as an efficient tissue adhesive to bind cells. Note that a supplementary procedure was used to avoid spurious effects due to other sources of adhesion on the cantilever or Petri dish. Therefore, a 0.1 g/dl BSA (Bovine Serum Albumin) solution was used to treat the cantilever with RBC just before the adhesion/separation experiment. This ensured that only the interaction between the cells was measured in the presence of dextran or fibrinogen.

The functionalised cantilever was then lowered onto a resting RBC in the Petri dish to capture it. After the cantilever was raised with the attached RBC, another RBC was chosen, and the cantilever was set on top of it and brought down until short contact was made at a setpoint force of 300 pN. Then, the cantilever was raised (5 µm/s) to separate the cells to determine the force-time signal. Based on this signal, the adhesion energy (the area under the curve in µJ) was obtained, and the interaction energy density ($\mu J/m^2$) was calculated by dividing the interaction energy by the RBC contact area (50 $\mu m^2$). This interaction energy was then plotted vs. the macromolecular concentration (fibrinogen or dextran).

Each experimental point represents 100 measurements for different cells obtained from different donors. Note that the RBC on the cantilever was changed for each concentration tested.

*Sedimentation measurements*

Absorption measurements at various concentrations of dextran and fibrinogen were performed to determine the sedimentation speed at 23°C. A cuvette was filled with a solution of washed RBCs at a density of 10 vol% (haematocrit) at various concentrations of dextran (MW = 70,000 kDa) or fibrinogen (both in powder form, Sigma Aldrich, Taufkirchen, Germany). The xenon lamp of the UV-Vis spectrophotometer Genesys 6 (Thermo Spectronic, Rochester, USA) was aligned 3 mm below the meniscus and adjusted to a wavelength of 940 nm. The cuvette (QS 170, quartz glass, Hellma analytics, Müllheim, Germany) had a layer thickness of 1 mm, a filling volume of 120 µl and dimensions of 35 x 12.5 x 12.5 (h x w x d in mm). Before each measurement, the cuvettes were cleaned with distilled water and ethanol.

*Microfluidic measurements*

Microfluidic devices were produced using the classic soft-lithography technique: PDMS was cast over a master obtained by photolithography of a SU8 photoresist over a silicon wafer and glued to a glass slide after plasma treatment. The microfluidic chip comprised 30 parallel channels with a rectangular cross-section of 8.5 µm in width, 4.5 µm in height and 5 cm in length. The channels were connected to a large inlet chamber where the blood sample was injected using a syringe pump. Images of the flowing cells at different locations along the channels were taken with a high-speed camera through an inverted microscope with a 10x objective lens. Nine channels were observed simultaneously. We ensured that the experimental conditions remained constant by taking several movie sequences (100 s each) at each location. Image processing allowed the detection of individual cells and clusters. Two cells were considered as part of the same cluster when their membrane-to-membrane distance was smaller than 0.6 times their length.

*Numerical method*

The surface to volume ratio (or the enclosed area A to the perimeter L ratio in two dimensions) is taken to be that of a human RBC; more precisely, $\nu = (A/\pi)/(L/2\pi)^2 = 0.65$. The fluid inside and outside of the cell model (CM) obeys the Stokes equations. Blood plasma viscosity is taken to be equal to that of water, whereas inside the CM, the viscosity is taken to be 5 times that of the plasma, a value that is consistent with available data on haemoglobin solutions. The velocity at the walls (mimicking blood vessels) is taken to be zero (the no-slip boundary condition). At the CM, the velocity is continuous, and the resulting hydrodynamic forces are balanced by the membrane force (a local, instantaneous mechanical equilibrium due to the weakness of inertial forces). The membrane force is composed of three contributions: (i) the bending force, (ii) tension forces that maintain local membrane incompressibility, and (iii) the net depletion force that has a repulsive part at short distances, taking into account the steric repulsion between cells (and possible electrostatic force), and an attractive long range force modelling the depletion mechanism, consistent with the classical model of Neu and Meiselman [25]. The depletion energy is chosen as $\phi = 4 \varepsilon [(\sigma/r)^{12} - (\sigma/r)^{6}]$. This expression has a minimum at $r = 2^{1/6} \sigma$, where the energy has the value $-\varepsilon$; these are the only two parameters specifying the interaction energy. The depletion force is simply the derivative of the energy $\phi$, and each given point on CM interacts with all other points belonging to any other membrane. The CM can resist the flow due to the membrane bending force, and the membrane is locally inextensible. Simulations were performed using the Green's function technique with an attractive long-range force between the vesicles modelling the depletion mechanism. Using Green's function techniques, we can express the velocity of each point of any given cell as an integral over all boundaries (the cell boundaries plus the wall boundaries). We used a specific Green's function (which is distinct from the classical Oseen tensor, albeit several preliminary results were obtained using the Oseen tensor), which offers the great advantage of automatically incorporating the presence of the boundaries (mimicking the blood vessels) in that we do not need to integrate over the boundaries where the stresses are not known a priori. This provides us with a powerful and precise computation. Technical details on the method and the numerical implementation will be given elsewhere

Literature

1. Fahraeus, R. The Influence of the rouleau formation of the erythrocytes on the rheology of the blood. *Acta Med. Scand.* **161**, 151-165 (1958).
2. Merrill, E.W., Cokelet, G.C., Britten, A., Wells, R.E. Non-Newtonian rheology of human blood - effect of fibrinogen deduced by "subtraction". *Circ Res.* **13**, 48-55 (1963).
3. Popel, A.S., Johnson, P.C. Microcirculation and hemorheology. *Annu. Rev. Fluid Mech.* **37**, 43–69 (2005).
4. Bishop, J.J., Popel, A.S., Intaglietta, M., Johnson, P.C. Rheological effects of red blood cell aggregation in the venous network: A review of recent studies. *Biorheology* **38**, 263-274 (2001).
5. Marton, Zs., Kesmarky, G., Vekasi, J., Cser, A., Russai, R., Horvath, B., Toth, K. Red blood cell aggregation measurements in whole blood and in fibrinogen solutions by different methods. *Clin. Hemorheol. Micro.* **24**, 75-83 (2001).
6. Maeda, N., Seike, M., Kume, S., Takaku, T. Shiga, T. Fibrinogen-induced erythrocyte aggregation: erythrocyte-binding site in the fibrinogen molecule. *Biochim. Biophys. Acta* **904**, 81-91 (1987).
7. Merrill, E.W., Gilliland, E.R., Lee, T.S., Salzmann, E.W. Blood rheology: Effect of fibrinogen deduced by addition. *Circ. Res.* **18**, 437–446 (1966).
8. Brooks, D. The effect of neutral polymers on the electrokinetic potential of cells and other charged particles: IV. Electrostatic effects in dextran mediated cellular interaction. *J. Colloid Interf. Sci.* **43**, 714–726 (1973).
9. Chien, S., Jan, K. Red cell aggregation by macromolecules: roles of surface adsorption and electrostatic repulsion. *J. Supramol. Str.* **1**, 385– 409 (1973).
10. Asakura, S., Oosawa, F. On interaction between two bodies immersed in a solution of macromolecules. *J. Chem. Phys.* **22**, 1255–1256 (1954).
11. Baeumler, H., Donath, E. Does dextran really significantly increase the surface potential of human red blood cells? *Stud. Biophys.* **120**,113–122 (1987).
12. Evans, E., Needham, D. Attraction between lipid bilayer membranes in concentrated solutions of nonadsorbing polymers: comparison of mean-field theory with measurements of adhesion energy. *Macromolecules* **21**,1822–1831 (1988).
13. Steffen, P., Verdier, C., Wagner, C. Quantification of depletion-induced adhesion of red blood cells. *Phys. Rev. Lett.* **110**, 018102 (2013).
14. Chien, S. Shear dependence of effective cell volume as a determinant of blood viscosity, *Science* **168**, 977-979 (1970).
15. Ziherl, P., Svetina, S. Flat and sigmoidally curved contact zones in vesicle-vesicle adhesion. *P. Natl Acad. Sci. USA* **104**, 761–765 (2006).
16. Jan, K.-M., Chien, S. Role of surface electric charge in red blood cell interaction. *J. Gen. Physiol.* **61**, 638-654 (1973).
17. Wells, R.E., Gawronski, T.H., Cox, P.J., Perera, R.D. Influence of fibrinogen on flow properties of erythrocyte suspensions. *Am. J. Physiol.* **207**, 1035-1040 (1964).
18. Chien, S., Usami, S., Taylor, H.M., Lundberg, J.L., Gregersen, M.I. Effects of hematocrit and plasma proteins on human blood rheology at low shear rates. *J. Appl. Physiol.* **21**, 81-87 (1966).
19. Chien, S., Usami, S., Dellenback, R.J., Gregersen, M.I., Nanninga, L.B., Guest, M.M. Blood viscosity: Influence of erythrocyte aggregation. *Science* **157**, 829-831 (1967).
20. Robertson, A.M., Sequeira, A., Kameneva, M.V. *Hemodynamical Flows: Modelling, Analysis and Simulation*, chapter Hemorheology, 63–120. Birkhaeuser, Basel (2008).

21. Baskurt, O.K., Temiz, A., Meiselman, H.J. Red blood cell aggregation in experimental sepsis. *J. Lab. Clin. Med.* **130**, 183-190 (1997).
22. Baskurt, O.K., Meiselman, H.J. Blood rheology and hemodynamics. *Semin. Thromb. Hemost.* **29**, 435–450 (2003).
23. Reinke, W., Gaethgens, P., Johnson, P.C. Blood viscosity in small tubes: effect of shear rate, aggregation, and sedimentation. *Am. J. Physiol.-Heart C.* **253**, H540-H547 (1987).
24. Cokelet, G.R., Goldsmith, H.L. Decreased hydrodynamic resistance in the two-phase flow of blood through small vertical tubes at low flow rates. *Circ. Res.* **68**, 1-17 (1991).
25. Popel, A.S., Johnson, P.C., Kameneva, M.V., Wild, M.A. Capacity for red blood cell aggregation is higher in athletic mammalian species than in sedentary species. *J. Appl. Physiol.* **77**, 1790–1794 (1994).
26. McWhirter, J.L., Noguchi, H., Gompper, G. Flow-induced clustering and alignment of vesicles and red blood cells in microcapillaries. *P. Natl Acad. Sci. USA* **106**, 6039–6043 (2009).
27. Tomaiuolo, G., Lanotte, L., Ghigliotti, G., Misbah, C., Guido, S. Red blood cell clustering in poiseuille microcapillary flow. *Phys. Fluids* **24**, 51903 (2012).
28. Fedosov, D.A., Pan, W., Caswell, B., Gompper, G., Karniadakis, G.E. Predicting human blood viscosity in silico. *Proc. Natl Acad. Sci. USA* **108**, 11772–11777 (2011).
29. Comeglio, P. et al. Blood clotting activation during normal pregnancy, *Thromb. Res.* **84**, 199-202 (1996).
30. Neu, B., Meiselman, H.J. Depletion-mediated red blood cell aggregation in polymer solutions. *Biophys. J.* **83**, 2482–2490 (2002).
31. Mchedlishvili, G., Gobejishvili, L., Beritashvili, N. Effect of intensified red blood cell aggregability on arterial pressure and mesenteric microcirculation, *Microvasc. Res.* **45**, 233-242 (1993).
32. Charansonney, O., Mouren, S., Dufaux, J., Duvelleroy, M., Vicaut, E. Red blood cell aggregation and blood viscosity in an isolated heart preparation, *Biorheology* **30**, 75-84 (1993).
33. Laschke, M.W., Vollmar, B., Menger, M.D. The dorsal skinfold chamber: window into the dynamic interaction of biomaterials with their surrounding host tissue. *Eur. Cells Mater.* **22**, 147-167 (2011).

**Acknowledgments**

This work was supported by the German Science Foundation research initative SFB1027 and the graduate school GRK 1276, by DFH/UFA (the German French University), by CNES (Centre National d'Etudes Spatiales) and by ESA (the European Space Agency).

**Author contributions**

C. W., C. M., T. P., G. C. and A.B. designed the research; M. B., D. F., L. K., M. L. and C. V. performed the experimental measurements; and O. A., H. S. and M. T. performed the numerical simulations. All authors discussed the results and helped in writing the paper.

**Author information**

Reprints and permission information is available at www.nature.com/reprints. The authors declare no competing financial interests. Readers are welcome to comment on the online version of the paper.

Figures

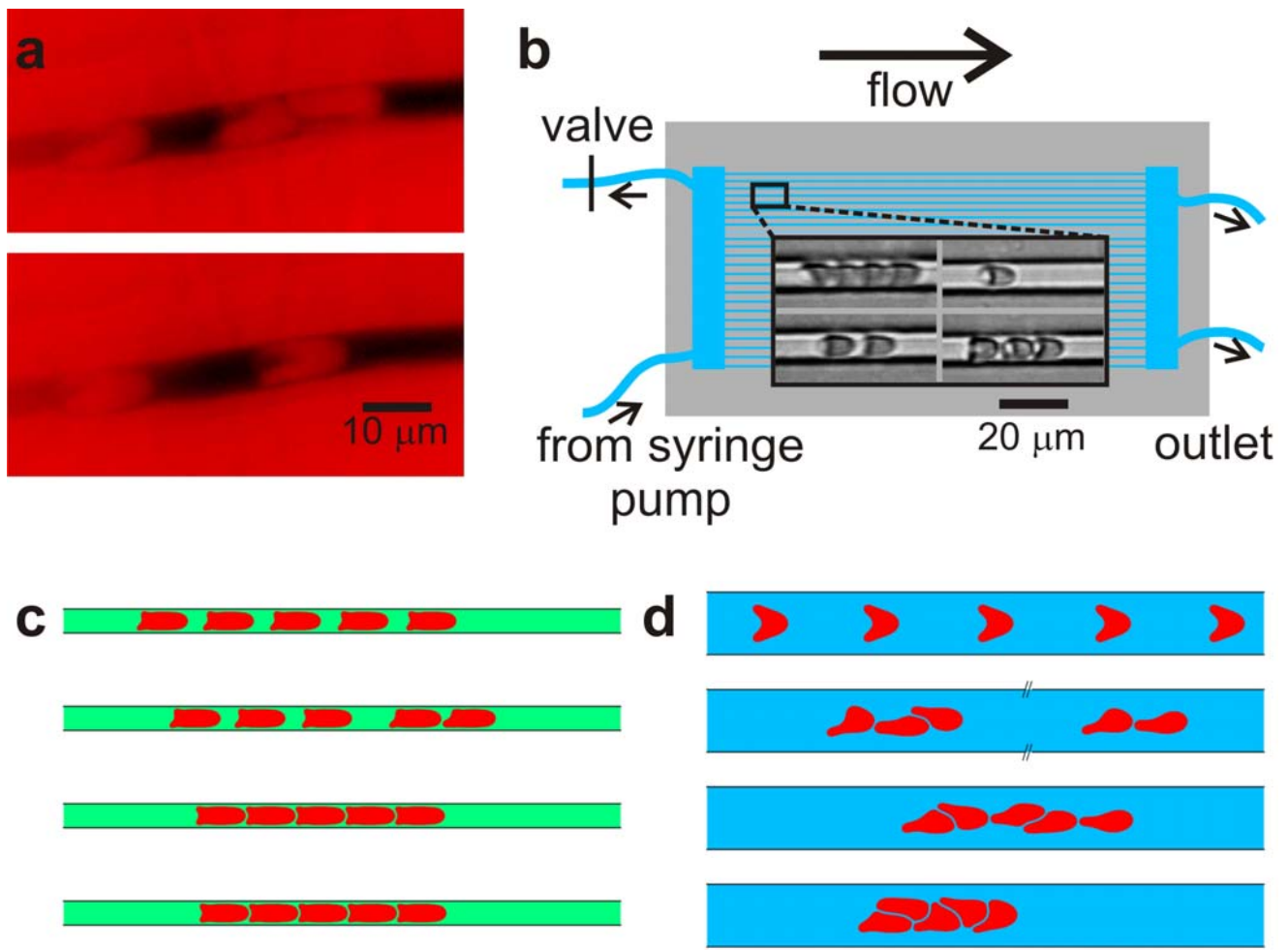

**Fig. 1**

a) A cluster of RBCs (above) and a single RBC (below) passing through a capillary of a mouse (see SM1 for an online video). The RBCs are imaged in a dorsal skinfold chamber using intravital fluorescence microscopy. To enhance the contrast, 0.05 ml 5% fluorescein-isothiocyanate (FITC)-labelled dextran was injected in vivo via the retrobulbary space. The effective dextran concentration in the vascular network is well below any significant effect of aggregation. The images are colourised. b) A schematic drawing of the microfluidic device and shadowgraph snapshots of clusters of RBCs in the microcapillaries. The microfluidic device includes 30 parallel channels of rectangular cross-section 8.5 μm (width) and 4.5 μm (height). The images are recorded using a high-speed camera, and the flow velocity and distribution of cells flowing through the observed section of the channels are determined (i.e., the numbers of single cells or clusters of different lengths (2, 3, 4, 5 or more cells) are counted). c) A snapshot of the steady-state numerical results regarding the effect of intercellular surface energy due to fibrinogen on RBC organisation in a channel with a width w = 4.5 μm. From top to bottom: $\varepsilon = 0$, $\varepsilon = 1.78$, $\varepsilon = 3.56$, and $\varepsilon = 4.89$ μJ/m² correspond to fibrinogen concentrations of 0, 1, 4 (still in the physiological range) and 6.5 mg/ml, respectively. d) The same as in c), but a channel of width (w) = 12 μm was used.

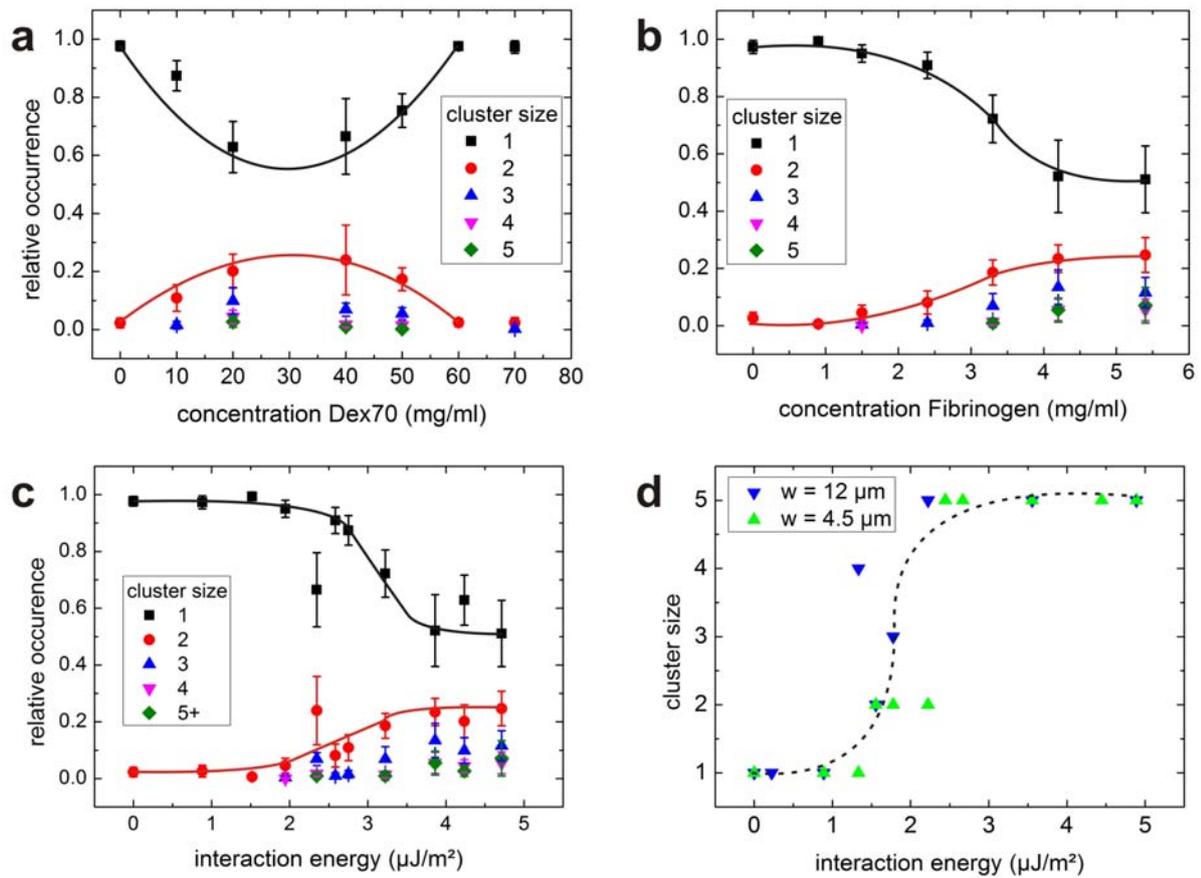

**Fig. 2**

a) Statistical distribution of differently sized RBC clusters at various dextran concentrations in a microfluidic device. At concentrations of 0 and greater than 60 mg/ml, no clusters were found. At 30 mg/dl, cell aggregation was so severe that clogging of the channels prevented systematic measurements. b) The same as in a) but using fibrinogen. Clustering increases with fibrinogen concentration. At concentrations greater than 5.5 mg/ml, the protein could not be fully dissolved; nevertheless, a stronger tendency to form aggregates and clog channels was observed. The lines are drawn as guides for the eye. c) Statistical distribution of differently sized clusters in the microfluidic device as a function of the interaction energy. d) Numerical results for the length of clusters at the steady state as a function of intercellular surface energy in channels of various widths.

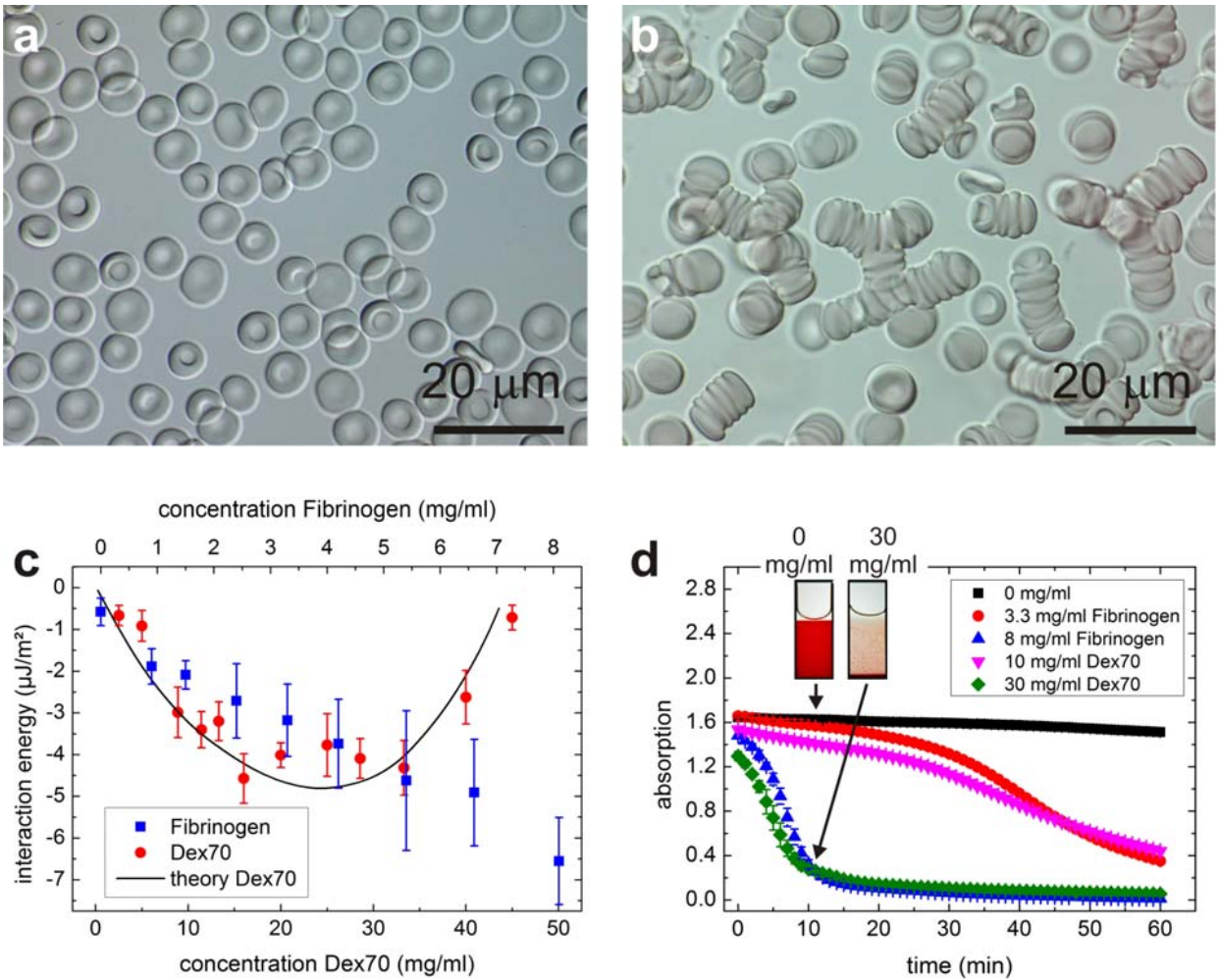

**Fig. 3**

An image of single RBCs a) in a buffer solution b) in rouleaux in a 20 mg/ml dextran solution in a Petri dish. c) Interaction energies between two RBCs based on single cell force spectroscopy (symbols) and the theoretical data from[25] (line). The experimental data obtained at various concentrations of dextran are from[11]. d) Absorption measurements at different concentrations of dextran and fibrinogen showing differences in sedimentation speed. A cuvette is filled with a suspension of washed RBCs at a density of 45 vol% (haematocrit) with various concentrations of dextran or fibrinogen. The beam from the UV-Vis spectrometer is aligned 3 mm below the meniscus. When the RBCs sediment, the opacity of the suspension decreases. Inset: Images of the cuvettes at 0 and 30 mg/ml dextran 11 minutes after filling. Aggregation of the cells leads to more rapid sedimentation.

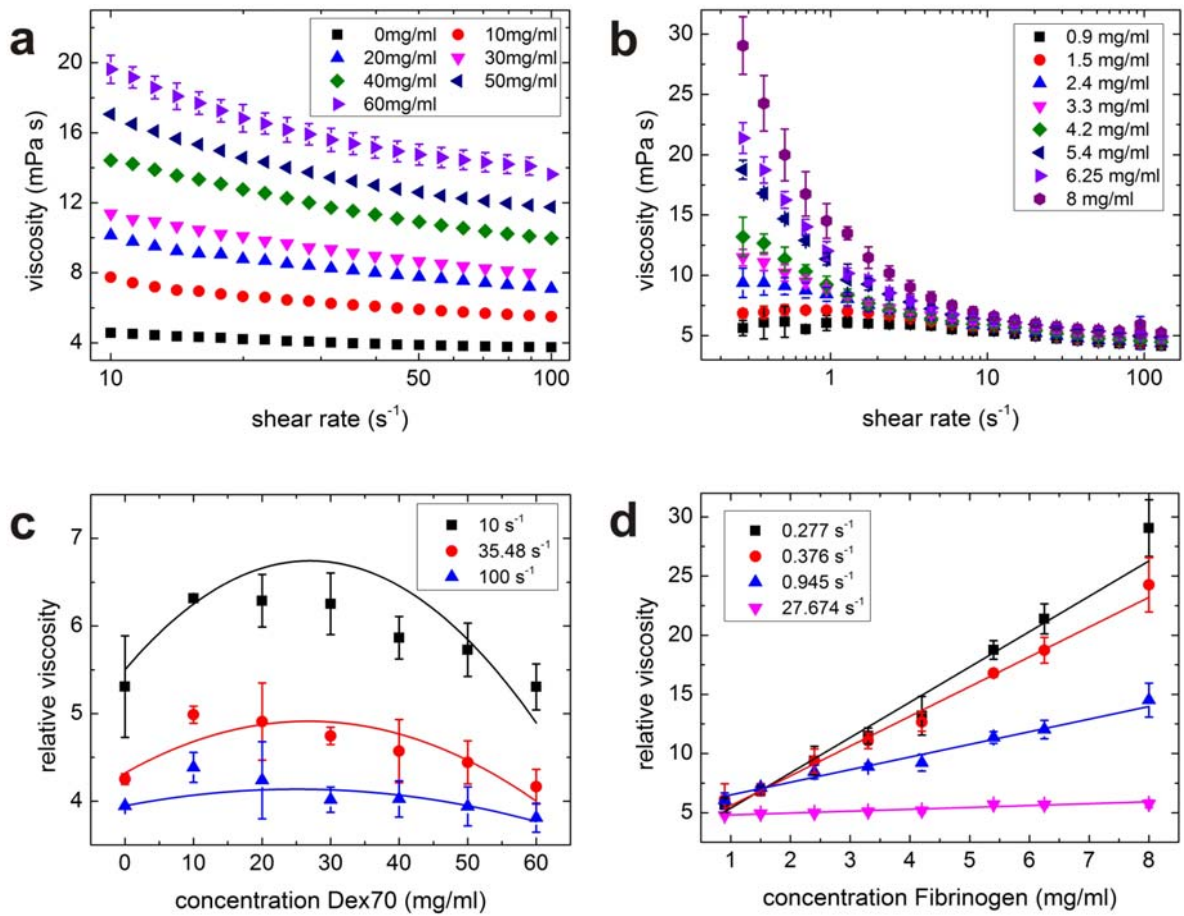

**Fig. 4**

a) Bulk viscosity of RBC suspensions at 45% haematocrit as a function of shear rate for various dextran concentrations. Error bars are only shown for the highest dextran concentration b) the same as a) but for different concentrations of fibrinogen. In both cases, the viscosity at low shear rates and the effect of shear thinning increases with macromolecule concentration. The viscosity of the solvent increases strongly with dextran concentration but is independent of fibrinogen concentration within the instrumental resolution. To separate the effect of aggregation in c) and d), the relative viscosity (the ratio of the viscosity of the RBC suspension and the viscosity of the solvent with dextran or fibrinogen only) is shown. Lines are drawn as guides for the eye.